\documentclass[aps,prl,twocolumn,superscriptaddress]{revtex4-1}  

\usepackage{amsmath}
\usepackage{amssymb}
\usepackage{graphicx}
\usepackage{epstopdf}
\usepackage{color}
\usepackage{cancel}

\begin{document}

\title{On-chip, photon-number-resolving, telecom-band detectors for scalable photonic information processing}

\author{Thomas Gerrits}
\affiliation{National Institute of Standards and Technology, Boulder, CO, 80305, USA}
\author{Nicholas Thomas-Peter}
\affiliation{Clarendon Laboratory, University of Oxford, Parks Road, Oxford, UK, OX1 3PU, United Kingdom}
\author{James C. Gates}
\affiliation{Optoelectronics Research Centre, University of Southampton, Highfield SO17 1BJ, United Kingdom}
\author{Adriana E. Lita}
\affiliation{National Institute of Standards and Technology, Boulder, CO, 80305, USA}
\author{Benjamin J. Metcalf}
\affiliation{Clarendon Laboratory, University of Oxford, Parks Road, Oxford, UK, OX1 3PU, United Kingdom}
\author{Brice Calkins}
\affiliation{National Institute of Standards and Technology, Boulder, CO, 80305, USA}
\author{Nathan A. Tomlin}
\affiliation{National Institute of Standards and Technology, Boulder, CO, 80305, USA}
\author{Anna E. Fox}
\affiliation{National Institute of Standards and Technology, Boulder, CO, 80305, USA}
\author{Ant\'ia Lamas Linares}
\affiliation{National Institute of Standards and Technology, Boulder, CO, 80305, USA}
\author{Justin B. Spring}
\affiliation{Clarendon Laboratory, University of Oxford, Parks Road, Oxford, UK, OX1 3PU, United Kingdom}
\author{Nathan K. Langford}
\affiliation{Clarendon Laboratory, University of Oxford, Parks Road, Oxford, UK, OX1 3PU, United Kingdom}
\author{Richard P. Mirin}
\affiliation{National Institute of Standards and Technology, Boulder, CO, 80305, USA}
\author{Peter G. R. Smith}
\affiliation{Optoelectronics Research Centre, University of Southampton, Highfield SO17 1BJ, United Kingdom}
\author{Ian A. Walmsley}
\affiliation{Clarendon Laboratory, University of Oxford, Parks Road, Oxford, UK, OX1 3PU, United Kingdom}
\author{Sae Woo Nam}
\affiliation{National Institute of Standards and Technology, Boulder, CO, 80305, USA}
\date{\today}

\begin{abstract}
Integration is currently the only feasible route towards scalable photonic quantum processing devices that are sufficiently complex to be genuinely useful in computing, metrology, and simulation.  Embedded on-chip detection will be critical to such devices.
We demonstrate an integrated photon-number resolving detector, operating in the telecom band at $1550$\,nm, employing an evanescently coupled design that allows it to be placed at arbitrary locations within a planar circuit.  Up to 5 photons are resolved in the guided optical mode via absorption from the evanescent field into a tungsten transition-edge sensor. The detection efficiency is $7.2 \pm 0.5\,\%$.  The polarization sensitivity of the detector is also demonstrated.  
Detailed modeling of device designs shows a clear and feasible route to reaching high detection efficiencies.
\newline
*Contribution of NIST, an agency of the U.S. Government, not subject to copyright
\end{abstract}

\maketitle

Photonics provides a promising path for building and using complex quantum systems for both exploring fundamental physics and delivering quantum-enhanced technologies in information processing, metrology, and communications.  Currently, the only feasible route towards sufficient complexity is integration, due to the high density of optical modes that can be contained within a single device and the extraordinary level of control that can be exercised over them.  Although much research has gone into developing integrated elements at telecom wavelengths for classical applications, their use in the quantum regime has been limited, in large part because of intrinsic inefficiencies in input coupling, detector coupling, and propagation.  The effect of these inefficiencies is to reduce or remove any quantum advantage attainable with a given device~\cite{VarnavaM2006lto, VarnavaM2008hgm, Politi:2008, Politi:2009, Thomas-PeterNL2010rqs, DattaA2010qmi, Thomas-PeterNL2011ips}.  Developing high-efficiency detectors that are compatible with these complex, high-density systems is therefore a critical enabling step for quantum photonics.

Current single-photon-sensitive detectors for telecom wavelengths include avalanche photodiodes (APDs)~\cite{Campbell2004}, superconducting nanowires~\cite{stevens2006}, and superconducting transition-edge sensors (TES)~\cite{MillerAJ03, LitaAE08}.  InGaAs APDs, the only commercially available telecom-band, single-photon-sensitive detectors suffer from low efficiencies and high dark-count rates, whereas nanowire detectors are extremely fast and can have relatively high quantum efficiencies compared to that of InGaAs APDs~\cite{Miki10, Fiore:11}.  Neither of these, however, has the ability to intrinsically resolve photon number.  By contrast, TESs, which function well across a very broad range of wavelengths, have been used to demonstrate the highest recorded detection efficiencies and number resolution, reaching 98\%~\cite{Fukuda:11} and can determine the photon number in pulses of light with up to 10 to 30 photons depending upon the design~\cite{MillerAJ03}.  All of these technologies typically employ designs in which the detected photons are normally incident on the detector.  In order to achieve high efficiencies in this situation, care must be taken to impedance-match the incident field to the detector in order to avoid reflections of the optical signal.  Moreover, normal incidence-detection schemes are intrinsically limited to monitoring the modes that emerge from the end facet of the device. By use of this detection scheme, inferring information about a quantum state or circuit element inside a device will only become more problematic as circuits move towards the complexities required to study effects beyond the scope of classical computational power~\cite{LaingA2010hoq, PeruzzoA2010mqi, Thomas-PeterNL2011ips}.

\begin{figure}
\begin{minipage}[b]{3.25in}
\includegraphics[width=3.25in]{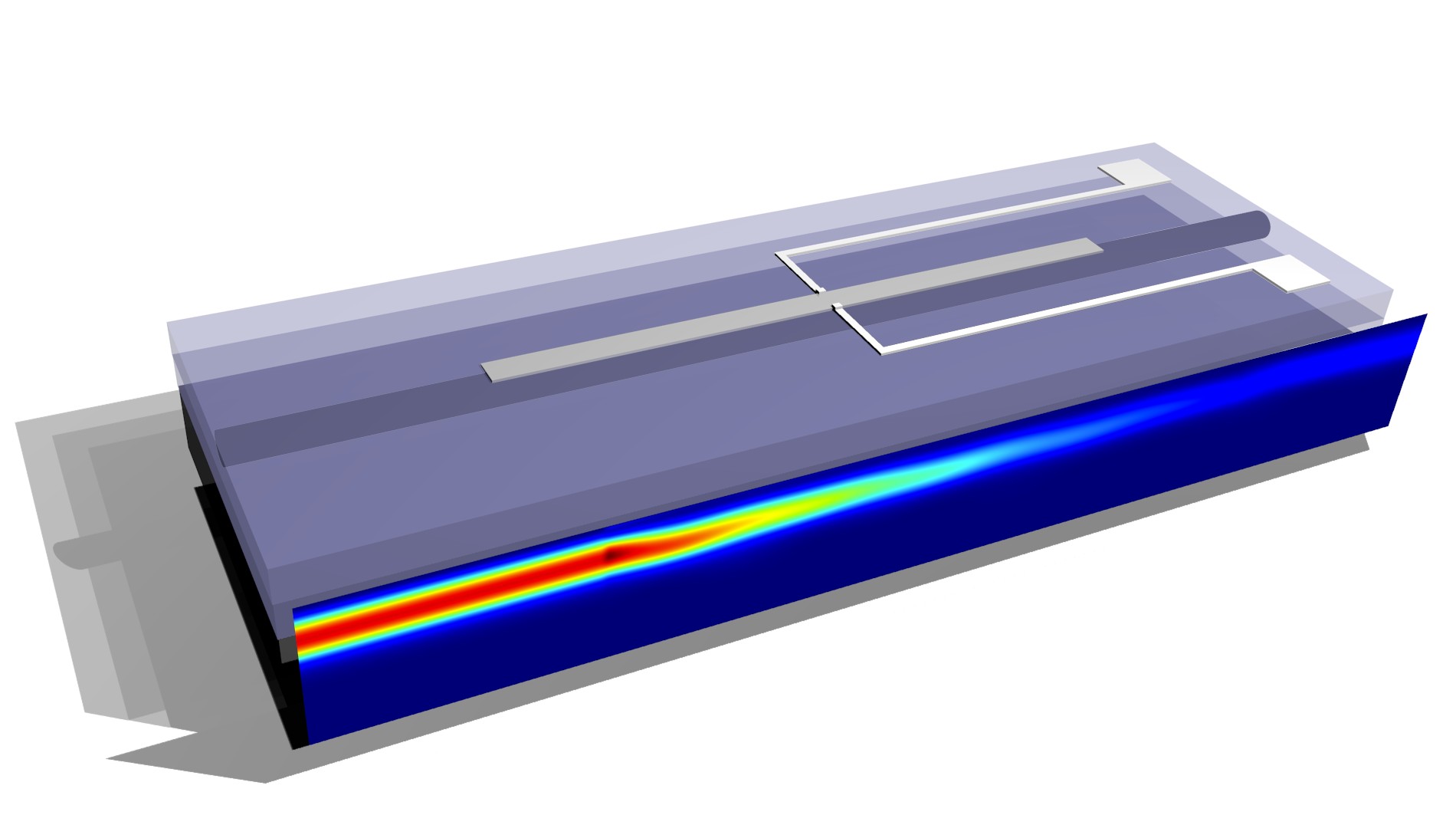}
(a) 
\vspace{0.1cm}
\end{minipage}

\begin{minipage}[t]{3.25in}
\includegraphics[width=3.25in]{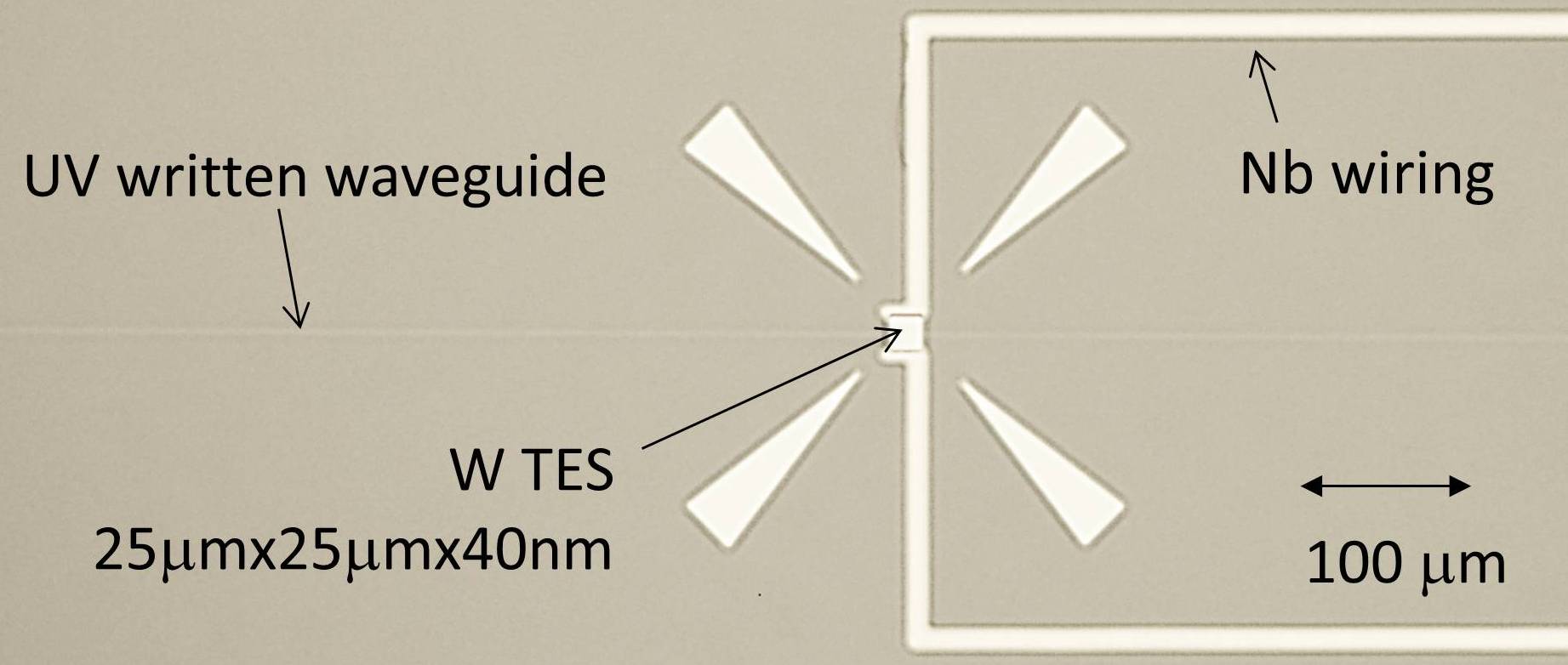}
(b)
\end{minipage}
\caption{(a) Schematic of the evanescently coupled photon counting detector. The detection layer is deposited on top of a UV-laser written Silica waveguide structure. The calculated internal mode profile can also be seen at the front of the device schematic. (b) Microscope image of the fabricated TES on the optical waveguide. The TES dimension is $25\,\mu \rm{m} \times 25\,\mu \rm{m} \times 40\,\rm{nm}$.  The tungsten TES bisects the waveguide in the center. The niobium wiring attaches top and bottom of the TES and exits off the chip to the right. Additional arrows simply serve as alignment marks and guide.}
\label{fig:conceptAndImage}
\end{figure}

In this paper we demonstrate the operation of a new concept for broadband, efficient, single-photon detection: Evanescently Coupled Photon Counting Detectors (ECPCDs), by merging two well-established technologies: photonic circuits and photon-number-resolving TESs.  ECPCDs display the low noise and single-photon sensitivity required to operate in the quantum regime and, due to the evanescent coupling concept, are integration compatible, as they target a single guided mode at an embedded location within the circuit and maintain a fixed alignment.  The photon absorber, in this case a tungsten TES, is placed in the evanescent field of the guided mode of a waveguide passing underneath the detector, as shown in Fig.~\ref{fig:conceptAndImage}(a).  As the mode propagates through the detection region, it is coupled continuously into the detector via absorption.  The continuous, evanescent coupling and the confinement of the waveguide mode allow potential for extended interaction regions, providing a straightforward way to increase efficiency even when the coupling is weak, without limiting the detector acceptance bandwidth.  A key feature of such a configuration is that the weak coupling allows use of a detector design that does not significantly modify the effective index and spatial profile of the guided mode, which minimizes reflections at the beginning of the detection region.  The detector is therefore almost fully impedance-matched.  Finally, the lithographic construction makes this concept compatible with fabrication techniques for conventional telecom planar lightwave circuits, enabling sufficiently high detector densities to be useful with complex circuit designs.

TESs are sensitive detectors that use the electrical/thermal properties of a superconducting thin film to distinguish the energy of a discrete number of photons.  By positioning the material at its superconducting transition temperature $T_{\rm c}$ via voltage biasing, absorbed photons create a temperature change that can be measured via the resulting change in  resistance~\cite{irwin:1998}.  To resolve this temperature change, the device's heat capacity must be carefully specified, which in turn limits the volume of superconducting material that may be used.  The choice of detector geometry is therefore critical to ensure good energy resolution and high-efficiency absorption.  In these first experiments to test the concept of a TES-based ECPCD, we chose detector dimensions of $25\,\rm{\mu m}\times25\,\rm{\mu m}\times40\,nm$~\cite{MillerAJ03}, which has been shown to provide number resolution of tens of photons, while still exhibiting an easily accessible superconducting transition temperature.

\begin{figure}
\begin{minipage}{0.49\linewidth}
\centering
(a)
\vspace{-0.1cm}
\end{minipage}
\begin{minipage}{0.49\linewidth}
(b)
\vspace{-0.1cm}
\end{minipage}
\includegraphics[width=\linewidth]{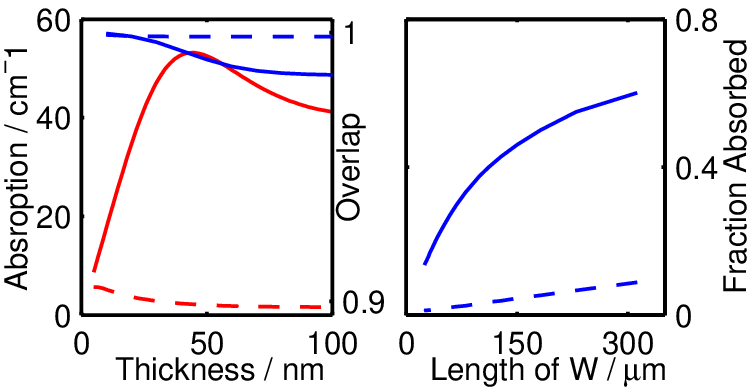}
\begin{minipage}{0.49\linewidth}
\includegraphics{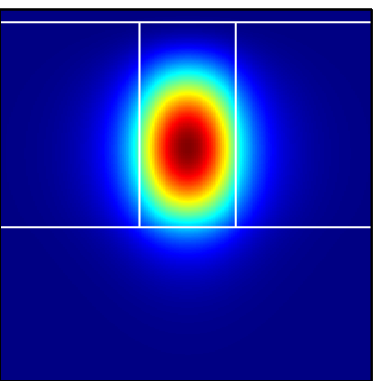}
(c)
\end{minipage}
\begin{minipage}{0.49\linewidth}
\includegraphics{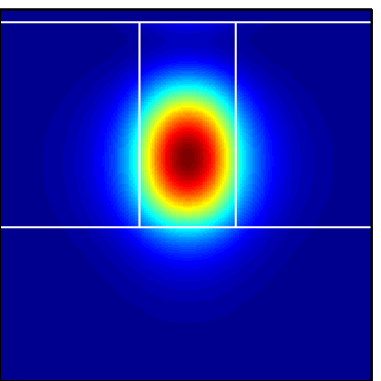}
(d)
\end{minipage}
\caption{(a) Plot showing how the overlap between the mode supported by the regions with and without the detector (blue) and the absorption coefficient ($\alpha/{\rm cm^{-1}}$, red) of both TE (dashed) and TM (solid) modes varies as the thickness of the layer is altered for a $25\,{\rm \mu m}$ wide detector. The maximum absorption coefficient for the TM mode occurs at around 45 nm and the preferential absorption of the TM-like mode is clear. (b) Plot showing how the absorption for the TE (dashed) TM (solid) mode changes as the length of the detector is increased while keeping the volume constant at $25\,\mu {\rm m}^3$. (c) The modeled intensity distribution of the TM mode in the region without the detector.  (d) The modeled intensity distribution of the TM mode in the region with the detector for a $40\,{\rm nm}$ thickness.  A slight distortion in the vertical direction can be seen along with a small amount of intensity near the detector layer.}
\label{fig:thicknessOverlapLength}
\end{figure}

Figure~\ref{fig:thicknessOverlapLength}(a) shows the calculated absorption coefficient for the TE and TM modes of the waveguide as the thickness of a tungsten layer $25\,{\rm\mu m}$ wide is increased.  This calculated absorption coefficient includes both the absorption in the tungsten TES and leakage from the propagating mode, as our modeling does not discriminate between them.  A clear peak in the absorption coefficient for the TM mode is visible at around 45\,nm thickness.  Initially, the absorption coefficient increases, because of the increase in volume of absorbing material interacting with the propagating field, but as the layer thickness increases beyond the skin depth of tungsten and the field no longer penetrates through the layer, the guided mode begins to be pushed away from the tungsten, and the total flow of energy into the tungsten starts to be reduced.  Also shown is the overlap of the TE and TM modes underneath the tungsten with those in the guide with no overlayer of tungsten, clearly demonstrating how little the spatial mode is perturbed by the presence of the detector.  To test our model, a sample was fabricated with a $100\,\mu$m long, $40 \rm{\,nm}$ thick strip of tungsten across the waveguide. The measurements were done by use of a method described in ref.~\cite{Rogers:10}. The absorption coefficients for the TE and TM modes were measured to be $3 \pm 2\,{\rm cm^{-1}}$ and $55 \pm 2.4\,{\rm cm^{-1}}$, which is in good agreement with our predicted values of $2.3\,{\rm cm^{-1}}$ and $54.6\,{\rm cm^{-1}}$.  Figure~\ref{fig:thicknessOverlapLength}(b) shows the calculated absorption as the length and width of the detector are varied at 40\,nm thickness in order to maintain a constant volume of $25\rm{\,\mu m^3}$.  It can clearly be seen that absorption greater than 50\,\% can be reached in a single device of this volume by employing a longer, narrower geometry.  This could be further improved by either increasing the detector volume or fabricating a multiplexed array of devices.

For our test devices, the guiding structure was provided by a direct UV-written waveguide.  These were produced by focusing a continuous-wave, 244 nm wavelength laser into a $5\,\rm{\mu m}$ thick photosensitive layer of germanium-doped silica situated on top of an undoped silica layer and a silicon substrate.  There are two key features of this waveguide platform for our devices.  First, the top surface of the chip is smooth to less than 1\,nm and is intrinsically planarized, which is critical, since surface roughness can suppress the tungsten superconducting transition temperature, $T_{\rm c}$.  Indeed, the observed $T_{\rm c}$ of $\sim90\,\rm{mK}$ is in good agreement with the value we expect for a 40\,nm thick film of tungsten.  Second, the guided mode is well mode-matched to standard telecom fiber so that the device can be pigtailed with relatively low loss.  Two TES detectors were fabricated, one positioned directly over the waveguide (as shown in Fig~\ref{fig:conceptAndImage}(b)), and one $800\,\rm{\mu m}$ away from it (not shown).  The former is the main interest of this study, while the latter allowed the effects of scattered light to be observed. 

\begin{figure}
\includegraphics[width=3.25in]{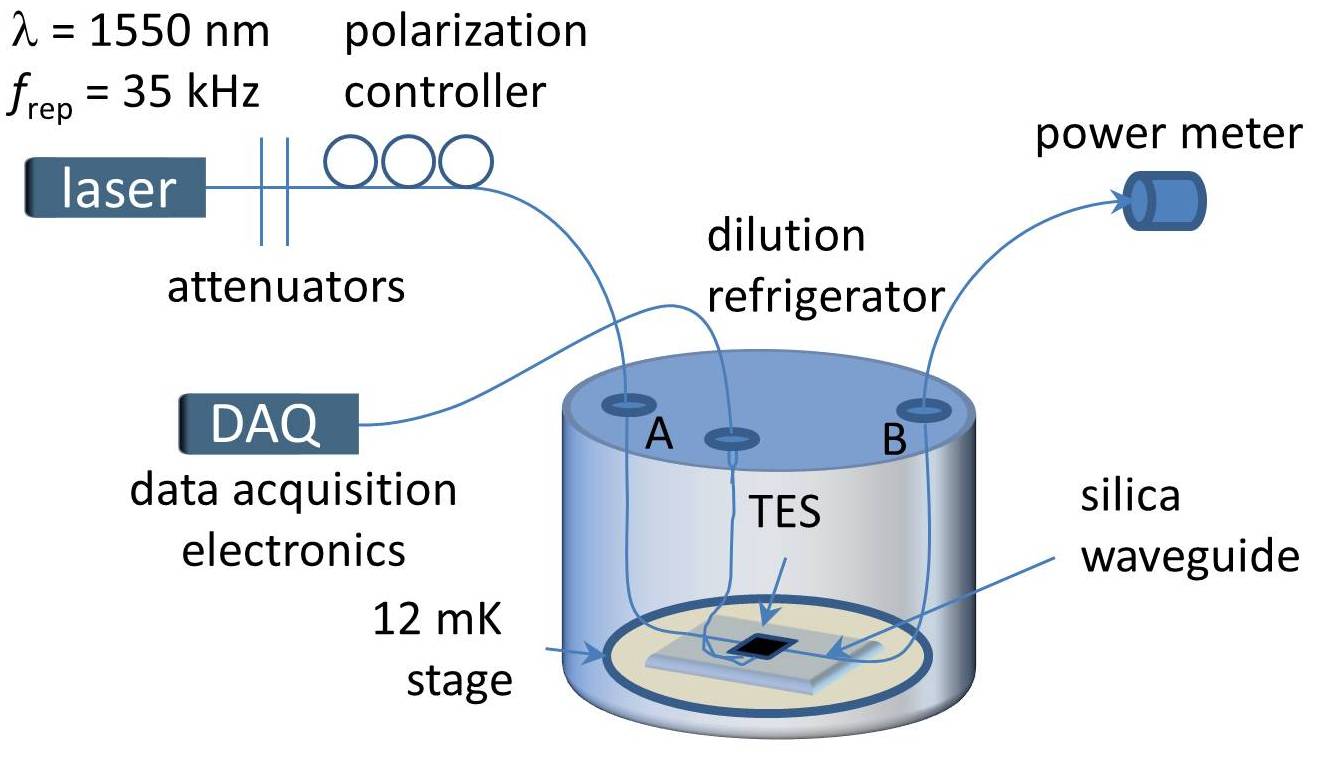}
\caption{Experimental scheme. A 35\,kHz pulsed laser at 1550\,nm delivers attenuated weak coherent state pulses to the waveguide device which is situated inside a dilution refrigerator. The input light's polarization can be modified via a fiber polarization controller. The waveguide device along with the TES is cooled to about 12\,mK. The TES is voltage biased and the electrical output is fed into a SQUID circuit (not shown). The SQUID output is amplified at room temperature and the signal is measured with data acquisition electronics.}
\label{fig:DR_fig}
\end{figure}

A schematic of the experimental setup used to characterize the detectors is shown in Fig~\ref{fig:DR_fig}.  The device was pigtailed and glued at both input and output before being cooled to $\sim12\,\rm{mK}$ in a commercial dilution refrigerator, significantly below the $T_{\rm c}$ for tungsten.  Pigtailing both end-facets of the device allowed us to test the loss in both directions and, by adjusting the input polarization, measure a total device throughput maximum of $19.1\pm 0.4\,\%$ and minimum of $17.7\pm 0.4\,\%$, corresponding to TE and TM modes, respectively.  The detectors were voltage biased within their superconducting transition region and the electrical output of each fed into a SQUID circuit (not shown).  An electronically pulsed 1550\,nm laser diode producing coherent states with 10\,ns duration at $\sim35\,\rm{kHz}$ was attenuated to a mean photon number of $\langle n \rangle = 28.5 \pm 0.8$ per pulse by use of two calibrated fiber attenuators~\cite{SOM}.  
  
\begin{figure}
\includegraphics[width=3.25in]{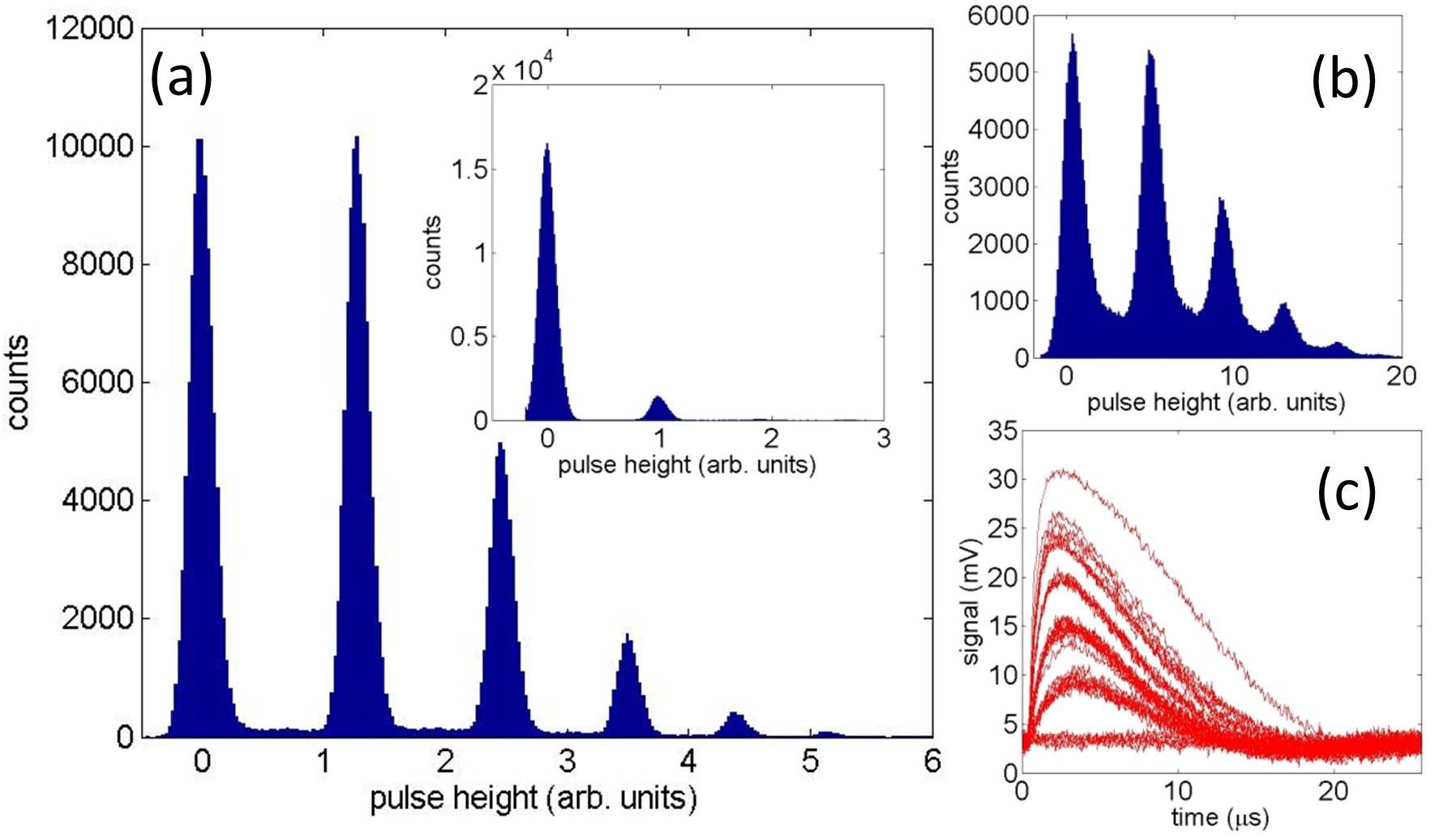}
\caption{Experimental results. (a) photon pulse height distribution for a measured coherent state with $\langle n \rangle = 0.986$ for the optimal TM polarization ($\eta_{det} = 7.2\,\%$). The inset shows the photon pulse-height distribution for the anti-optimal TE polarization. The measured mean photon number was $0.086$ ($\eta_{det} = 0.65\,\%$). (b) photon pulse height distribution for the TES not evanescently coupled to the waveguide ($\langle n \rangle = 1.03$; $\eta_{det} = 0.0056\,\%$). (c) electrical TES output traces for different numbers of photons in the weak laser pulse; the photon number resolving capability is clearly visible here. The TES recovery time is about $10\,\rm \mu s$ due to its low transition temperature. Note that this detector was optimized for its photon-number-resolving capability and the evanescent coupling strength, not for its recovery time.}
\label{fig:data}
\end{figure}

The raw output pulses of the central TES are shown overlaid on top of each other in Fig.~\ref{fig:data}(c), and a histogram of pulse heights is shown in Fig.~\ref{fig:data}(a) for the TM mode (the inset of Fig.~\ref{fig:data}(a) for the TE mode).  The clearly observable separation between pulse heights (the clearly resolved peaks in the histogram) demonstrates how well this detector can resolve photon number and shows that simple thresholding electronics are sufficient to determine the absorbed photon number.  The measured mean photon number per pulse was $0.986\pm0.02$ for the TM mode, found by optimizing the detector response while adjusting the input polarization. A mean photon number of $0.086\pm0.008$ was found for the TE mode by adjusting the input polarization to the smallest detector response. We measured the photon number statistics and determined the mean photon number per pulse for both propagation directions to allow us to account for the effects of coupling and propagation loss~\cite{SOM}. This method yielded an absolute detection efficiency of $7.2\pm 0.5\,\%$ and $0.65\pm 0.05\,\%$ for the TM and TE modes, respectively.  We compare this with our simulations, which predict absorption coefficients of 13.2\,\% and 1.2\,\% for our detector geometry. One possible explanation for this discrepancy is the fact that the model prediction does not discriminate between actual absorption of the tungsten TES and possible leakage of light from the waveguide structure, while the measurement provides only the absorption in the TES. In addition, fiber-splice losses in our system may also contribute to a measured detection efficiency a few percent lower than predicted~\cite{MillerAJ2011}.

Figure~\ref{fig:data}(b) shows a pulse-peak histogram for the reference detector after the input mean photon number was increased nominally by 32\,dB and the input polarization was adjusted to obtain the maximum measured mean photon number.  This number was $1.03\pm0.02$, implying a detection efficiency of $0.0056\,\%$, which we expect for a detector that measures only the scattered photon contribution.  From the scattered light signal we estimate that this contribution to the central TES detection signal is $\sim 8\times10^{-4}$ per pulse.  The main peaks corresponding to direct photon number absorptions are still present, however there are also intermediate pulse heights that contribute to an exponential tail on what is usually a Gaussian peak.  We have evidence that these intermediate pulse heights are caused by a parasitic absorption of photons in the silica itself. An absorbed photon will create a cloud of hot electrons that couple to the silica phonon system. A part of this heat diffuses into the tungsten, causing a partial warm-up, {\it i.e.}, an energy collection efficiency lower than in the case of direct absorption~\cite{cabrera:1998}. The effect is less pronounced in the central TES, as the majority of the photon absorption is due to evanescent coupling into the tungsten directly. We also observe a very small peak at half the photon energy, a possible indication of photon absorption in the niobium wiring~\cite{cabrera:1998}. We expect these effects to be suppressed when more optimized detector geometries with higher efficiency are used in the future.

In this paper we have realized the concept of an evanescently coupled photon-counting detector and demonstrated its operational feasibility by constructing a waveguide-based transition edge sensor, the first implementation of an on-chip, truly photon number-resolving-detector.  We have shown a clear and realistic route to high-efficiency detection using this scheme by engineering the aspect ratio of the detector.  This wholly integrated solution for detection will be a key component of high efficiency integrated devices functioning in the quantum regime.  Such devices will be necessary in order to demonstrate technologies that show a palpable enhancement in functionality over their classical counterparts.

\begin{acknowledgements}
T.G. thanks Aaron J. Miller and M. Scott Bradley for discussions during the preparation of this manuscript.
\end{acknowledgements}

%\bibliography{arXiv}

\clearpage

%%%%%%

\section{Supporting Online Material}

The purpose of this supporting material is to provide technical details of the fabrication process, the experimental setup, the determination of the transition edge sensor (TES) detection efficiency and our modeling.

\subsection{TES background}

TESs are sensitive detectors that use the electrical/thermal properties of a superconducting thin film to distinguish the energy of a discrete number of photons. Superconducting metals exhibit a resistance change from a high-temperature value of several ohms to negligible material resistance values at temperatures below the critical transition temperature, $T_{c}$. Materials used to make detectors sensitive to optical photon energy have transition temperatures on the order of $100\,{\rm mK}$ with transition widths as narrow as $1$\, mK. By thermally positioning the metal in its superconducting transition via voltage biasing, single absorbed photons create a detectable resistance change~\cite{irwin:1998}. 
By use of this technique, materials such as tungsten or titanium can be fabricated into microcalorimeters sensitive enough to resolve the absorption of a single optical photon. These detectors have advantages over other types of single-photon detectors, making them ideal for integration into quantum optical systems. Efficiencies have been reported as high as $95\,\%$ at $1550$\,nm and they have a wavelength range exhibiting high quantum efficiency from the visible to the IR, making them more efficient and wavelength-expansive than any silicon or InGaAs detector~\cite{LitaAE08, HadfieldRH07}. Additionally, TES detectors exhibit a thermal recovery time as short as $1\,{\rm \mu s}$~\cite{HadfieldRH07}, and arrival-time jitter of less than $100$\,ns FWHM. Unlike most conventional high temperature photodetectors, TES detectors are number-resolving, which means these detectors can distinguish the energy correlated to the absorption of not only a single photon (“click detector”), but energy correlated to the absorption of several photons. To operate a TES, the bath temperature of the whole system is held below the critical temperature of the TES and a voltage bias is applied to the TES itself. Joule heating produced by this voltage bias is carefully controlled to increase the TES temperature in order to position it within its sharp transition region. The TES will remain in a steady-state at this temperature until a photon is absorbed.  In a properly thermally engineered TES, the absorption of a photon will result in an increase of the TES temperature, with a corresponding increase in the TES resistance and therefore a decrease in the current. Under constant-voltage bias, the drop in the TES current is proportional to the absorbed energy, and this current change is measured by use of DC-superconducting quantum interference device (SQUID) amplifier circuits.

\begin{figure}
\includegraphics[width=3.25in]{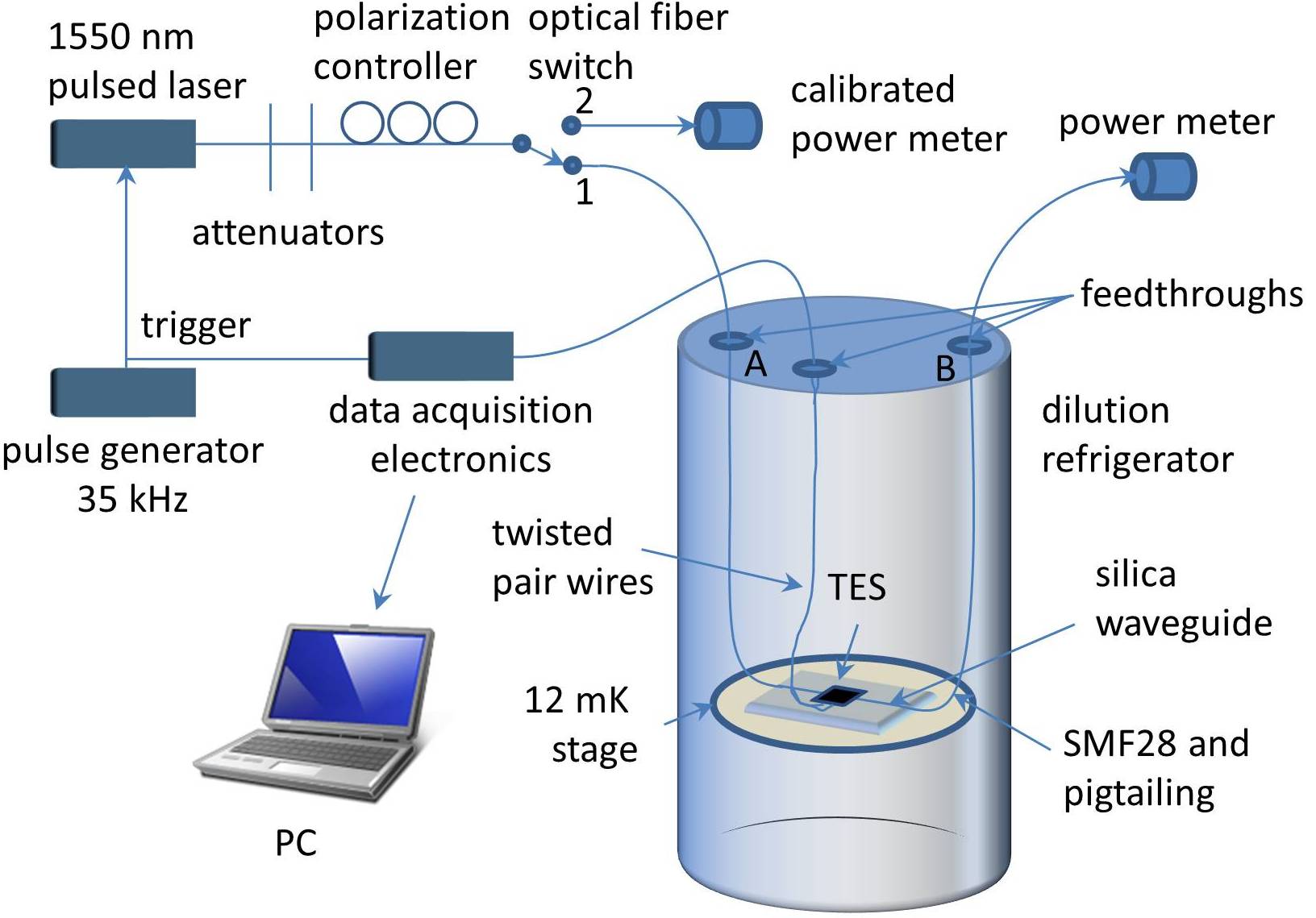}
\caption{Experimental setup for device testing. The TES and waveguide were held at a temperature of about $12$\,mK inside a commercial dilution refrigerator (DR). The output of a pulsed laser is attenuated to the single-photon level. The attenuated coherent light pulses are delivered to the device via a commercial telecom optical fiber. Calibration of the input mean photon number was done by use of a calibrated power meter to measure the average input power and the knowledge of the attenuator values, which were determined before each measurement. The fiber optical switch allowed calibration of the input signal before each measurement. The electrical signal from the TES detector was delivered to a SQUID circuit inside the DR (not shown) and subsequently amplified with a $\times\,100$ amplification stage outside the DR. The amplified signal containing the photon traces was recorded with data-acquisition electronics and post-processed to identify the pulse height of each recorded photon event trace.}
\label{DR_fig}
\end{figure}

\subsection{Waveguide and TES fabrication} 

An important advantage of integrating TES detectors into quantum optical systems is their ability to be implemented onto a waveguide structure. The waveguide structure used in this work was written by use of a UV femtosecond laser writing technique designed to alter the index of a silica core layer on top of a $17\,{\rm \mu m}$ thermal silicon oxide cladding layer. No top cladding was fabricated, to maximize the evanescent coupling to the TES. The planar core/cladding layer refractive index contrast is 0.6\,\% with core layer thickness of $5.5\,\mu {\rm m}$. The UV written channel was Gaussian in profile with a contrast of 0.3\,\% and a width of $\sim 6\,\mu {\rm m}$. The surface roughness of the intrinsically planarized waveguide structure is typically less than 1\,nm. This allows deposition of the TES with only a thin layer of amorphous silicon underneath to relieve the thermal stress at cryogenic temperatures. This deposition process is equivalent to the process used for fiber-coupled TES used in earlier studies~\cite{LitaAE08, LitaAE2010sts}. To fabricate the TES detector, a $40\,{\rm nm}$ thin film of tungsten was DC sputtered onto a thin layer of amorphous silicon. The tungsten layer was patterned and etched such that a $25\,{\rm \mu m}  \times 25\,{\rm \mu m}$ pad remained directly in contact with the core layer of the silica waveguide. Wiring to the tungsten was established by use of niobium, which was sputtered and lifted off.

\subsection{TES fabrication and efficiency calibration} 

Figure \ref{DR_fig} shows the experimental schematics for measuring weak coherent pulses of light with the Evanescently Coupled Photon Counting Detector (ECPCD). We use a dilution refrigerator (DR) to cool the TES below its transition temperature of $\sim\,90\,{\rm mK}$. A pulsed laser diode (driven with a pulse generator) produced coherent state pulses at a wavelength of $1550\,{\rm nm}$ with a temporal width of $10$\,ns and a repetition rate of $\sim 35\,{\rm kHz}$. The laser pulses were sent through an array of two fiber attenuators, a polarization controller, and an optical fiber switch, either to measure the output power directly with a calibrated power meter or to send the attenuated laser pulses toward the waveguide chip. We calibrated the attenuators before each measurement by individually measuring the un-attenuated and attenuated average laser power for each of the two attenuator settings~\cite{MillerAJ2011}. Typically we attenuated the laser pulses by about $60\,{\rm dB}$ to reach a mean photon number ($\langle n \rangle$) of about 30 at the input fiber after the optical switch. Before the optical switch we used a fiber polarization controller to modify the input polarization into the waveguide chip. The electrical output of the TES was fed into a SQUID circuit and amplified with a $\times\,100$ amplifier outside the dilution refrigerator. This signal was recorded with data-acquisition electronics and post-processed with a PC to determine the pulse height of each of the individual photon traces~\cite{AlbertoD2010,LitaAE08}. We can also measure the overall transmission of the waveguide by measuring the un-attenuated power with a second power meter at the output port of the waveguide. Measuring the TES response and transmitted power along both possible directions of the optical waveguide ($A \rightarrow B$ or $B \rightarrow A$), allowed determination of the coupling losses up to the TES and the TES detection efficiency. 

\begin{figure}
\includegraphics[width=3.25in]{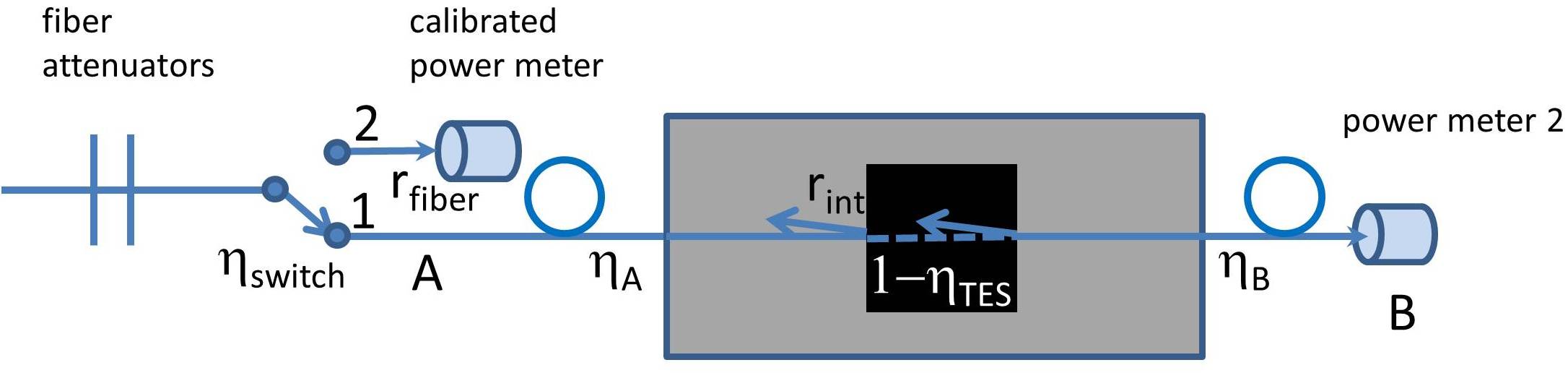}
\caption{Experimental scheme for measuring the coupling efficiency and detection efficiency of the TES detectors. Two power meters are used to measure the overall transmission of the fiber-pigtail-waveguide-TES array. We measure in both possible propagation directions. $\eta_{\rm switch}$ is the optical fiber switch efficiency, $r_{\rm fiber}$ is the silica-air reflection at the fiber end, $\eta_{A}$ and $\eta_{B}$ are the fiber, pigtail and waveguide transmissions up to the TES detector. For the full characterization of the TES detection efficiency we include the reflection at the waveguide-TES interface due to impedance mismatches. $\eta_{\rm TES}$ is the TES detection efficiency.}
\label{det_scheme_fig}
\end{figure}

Figure \ref{det_scheme_fig} shows the schematic for measuring the detection efficiency of the ECPCD. We used two power meters to measure the overall transmission along both possible propagation directions ($T_{AB}$ or $T_{BA}$). After calibration of the fiber attenuators we directed the attenuated laser pulses towards the waveguide chip. Therefore, we need to determine the switch imbalance between the two output ports. We model this imbalance equivalent to the weak laser pulse propagating through a beam splitter with transmissivity $\eta_{\rm switch}$. We measured $\eta_{\rm switch} = 91.7 \pm 1.0\,\%$. Our calibrated power meter is a free-space power meter. Therefore, we need to expect a silica-air reflection off the end of the optical fiber. This reflection ($r_{\rm fiber}$) is $3.60\,\%$, and the waveguide input power needs to be corrected by this amount.
The corrected input power ($P_{\rm in}$) is:
\begin{equation}
P_{\rm in} = P_{\rm pm} \times \frac{\eta_{\rm switch}}{1-r_{\rm fiber}},
\label{eq:trans2a}
\end{equation}
where $P_{\rm pm}$ is the measured power at the calibrated power meter.
Also, we compared the second power meter (power meter 2) to the calibrated power meter for the purpose of the waveguide transmission measurements. This gave a correction factor ($\gamma$), correcting for the readout offset of power meter 2. We measured $\gamma = 0.9995$. The overall transmission is equal to:
\begin{eqnarray}
T&=&T_{AB}=T_{BA} \nonumber \\
&=& \eta_{A} \times (1-r_{\rm int}) \times (1-\eta_{\rm TES}) \times \eta_{B}, \label{eq:trans1}
\end{eqnarray}
where $\eta_A$ and $\eta_B$ are the overall fiber-pigtail-waveguide transmissions up to the TES detector, $r_{int}$ is the reflection at the waveguide-TES interface, and $\eta_{\rm TES}$ is the TES detection efficiency including the second reflection at the TES-waveguide interface along with the subsequent reflections and absorption. It is reasonable to assume that both $\eta_A$ and $\eta_B$ are independent of the direction of the light propagation, as we measured an overall transmission $T_{AB} = 17.74\,\%$ and $T_{BA} = 17.65\,\%$ for each propagation direction, respectively. The difference between these two values is less than the measurement uncertainty. We measure the mean photon number of the input state that propagated from point $A$ or $B$ to the TES. Hence we can infer the overall system detection efficiencies $\eta'_A$ and $\eta'_B$￼ after the optical fiber switch, including the TES efficiency:
\begin{equation}
\eta'_{A/B} = \eta_{A/B} \times (1-r_{int}) \times \eta_{\rm TES}, \label{eq:trans2}
\end{equation} 
$\eta'_{A/B}$ are determined by use of the calibration routine described in~\cite{MillerAJ2011}. Solving for $\eta_{\rm TES}$ and applying the result to eqn.\,\ref{eq:trans1} and eqn.\,\ref{eq:trans2} gives:
\begin{eqnarray}
\eta_{\rm TES}&=&\frac{1}{2} \left (\sqrt{K\left ( K + 4\right )} -K\right) \nonumber \\
\eta_{A/B}&=&\frac{\eta'_{A/B}}{(1-r_{\rm int}) \times \eta_{\rm TES}}, \label{eq:trans3}
\end{eqnarray}
where $K=\eta'_A \times \eta'_B/\left [ T (1-r_{\rm int}) \right ]$. We used the mean of $T_{AB}$ and $T_{BA}$ ($T = 17.7 \pm 0.4\,\%$) and determined $\eta'_A￼= 2.9 \pm 0.2\,\%$ and $\eta'_B￼= 3.5 \pm 0.2\,\%$. From these measurements we cannot estimate or determine the waveguide-TES interface reflection $r_{\rm int}$. However, we can utilize the theoretical prediction as a basis for a good reflection estimate ($r_{\rm int(theory)} = 0.03\,\%$) and calculate the efficiencies assuming a reflectivity $r_{\rm int}$. Table\,\ref{tab:table1} shows the measurement results of the system detection efficiencies and the inferred efficiencies based on eqn.\,\ref{eq:trans3}.

\begin{table}
\caption{Extracted fiber-pigtail-waveguide transmission $\eta_{A/B}$ and TES detection efficiency $\eta_{\rm TES}$.}
\label{tab:table1} 
\centerline{
\begin{tabular}{|c|c|c|c|}
\hline
$r_{\rm int}$ & $\eta_A$ & $\eta_B$ & $\eta_{\rm TES}$\\
\hline
0.03\,\% &  $39.8 \pm 4.4\,\%$ & $47.9 \pm 5.0$ & $7.2 \pm 0.4\,\%$\\
1.00\,\% &  $40.0 \pm 4.4\,\%$ & $48.2 \pm 5.0$ & $7.3 \pm 0.4\,\%$\\
\hline
\end{tabular}
}
\end{table}

The TES detection efficiency is $\eta_{\rm TES} = 7.2 \pm 0.4\,\%$ for a theoretical reflectivity ($r_{\rm int}$) of $0.03\,\%$ and $\eta_{\rm TES} = 7.3 \pm 0.4\,\%$ for $r_{\rm int} = 1.00\,\%$. The uncertainty in $\eta_{\rm TES}$ was determined by propagating the measurement uncertainties. The main contribution to the measurement uncertainty is the determination of the absolute input power, and the attenuator calibration and transmission measurements. The relative uncertainty for each of the power measurements was $0.5\,\%$, an expanded uncertainty equivalent to the $2\sigma$ interval of the power meter calibration. The systematic uncertainty due to the unknown reflectivity at the waveguide-TES interface is smaller than the uncertainty due to the calibration routine. Adding the systematic uncertainty to the measurement uncertainty and taking the average yields: $\eta_{\rm TES} = 7.2 \pm 0.5\,\%$, $\eta_A = 39.8 \pm 4.6\,\%$ and $\eta_B = 47.9 \pm 5.2\,\%$.

\subsection{Device Modeling} 
A commercial numerical mode-solver based on film mode matching is used to model our device and investigate suitable design parameters. There are two limitations to the efficiency of our detector once light has been coupled into it.  First, the presence of the TES tungsten detection layer affects the transverse profile of the mode supported by the waveguide underneath it.  The resulting mode mismatch between the propagation and detection regions can result in a back reflection that will not be absorbed by the detector. Second, the absorption of the detector is governed by the shape and size of the tungsten layer.  However, the volume of a single detector is constrained by the detection mechanism.  Since photons are resolved by a temperature change of the electronic substructure, the heat capacity of the detector must be small enough such that the temperature change due to a single absorbed photon can be resolved.  Throughout our device optimization we fix the volume of the detector to $40\,{\rm nm} \times 25\,{\rm \mu m}  \times 25\,{\rm \mu m}$, a volume close to that which has previously been demonstrated to achieve high photon-number resolution and single-photon sensitivity~\cite{MillerAJ03}, while still having an accessible superconducting transition temperature.

The eigenmode expansion method is used to propagate our modes through the device and obtain the overall scattering matrix, from which we deduce the absorption coefficients due to the detector layer. This approach takes account of any reflections and mode mismatches at the interfaces to give an accurate model of the absorption coefficients at the detector region.

When we calculate the absorption coefficient for the TE and TM detection region modes as a function of detector thickness, we observe a clear peak in the absorption coefficient at a thickness of $45$\,nm and strong preferential absorption of the TM mode. Since this is already close to the $40$\,nm, a detector thickness that has previously been shown to work for fiber coupled detectors, we continued with this design geometry for the initial tests.  The numerical overlap of the TE and TM modes with the propagation region modes is $98.4\,\%$ and $97.6\,\%$, respectively, and the reflection at the mis-matched interface is small, at only $0.01\,\%$ and $0.03\,\%$ for the TE and TM modes.  The final calculated total absorption is $13.2\,\%$ for the TM and $1.2\,\%$ for the TE mode. The ratio of absorption coefficients between the TM and TE modes therefore is $0.909$.  Even though the absolute values for the estimated absorption are higher than the measured detection efficiencies, due to the inability of our model to distinguish the tungsten absorption and the waveguide losses, the relative absorption of the TM and TE modes are in very good agreement with the experimental result ($0.913$).

\bibliography{arXiv}

\end{document}